\documentclass[twocolumn,showpacs,amsmath,amssymb,prl]{revtex4-1}
\usepackage{graphicx} 
\usepackage{epsfig}
\usepackage{dcolumn} 
\usepackage{color}
\usepackage{changes}
\newcommand{\beq}{\begin{equation}}
\newcommand{\eeq}{\end{equation}}
\newcommand{\beqa}{\begin{eqnarray}}
\newcommand{\eeqa}{\end{eqnarray}}

\newcommand{\kvec}{{\bf k}}

\newcommand{\qvec}{{\bf q}}

\newcommand{\rvec}{{\bf r}}

\begin{document}
\title{Dynamical charge density waves rule the phase diagram of cuprates}
\author{S. Caprara$^{1,2}$, C. Di Castro$^{1,2}$, G. Seibold$^3$, and M. Grilli$^{1,2}$}
\affiliation{$^1$ Dipartimento di Fisica, Universit\`a di 
Roma ``La Sapienza'', P.$^{le}$ Aldo Moro 5, 00185 Roma, Italy \\
$^2$ ISC-CNR and Consorzio Nazionale Interuniversitario per le Scienze Fisiche della 
Materia, Unit\`a di Roma ``Sapienza''\\
$^3$ Institut f\"ur Physik, BTU Cottbus-Senftenberg - PBox 101344, D-03013 Cottbus, Germany}
\begin{abstract}
In the last few years charge density waves (CDWs) have been ubiquitously observed in high-temperature 
superconducting cuprates and are now the most investigated among the competing orders in the still hot debate 
on these systems. 
A wealth of new experimental data raise several 
fundamental issues that challenge the various theoretical proposals. Here, we account for the complex 
experimental temperature vs. doping phase diagram and we provide a coherent scenario explaining why
different CDW onset curves are  observed by different experimental probes and seem to extrapolate 
at zero temperature into seemingly different quantum critical points (QCPs) in the intermediate and overdoped region.
We also account for the pseudogap and its onset temperature $T^*(p)$ on the basis of dynamically fluctuating CDWs. 
The nearly singular anisotropic scattering mediated by these fluctuations also
account for the rapid changes of the Hall number seen in experiments and provides the first necessary step for
a possible Fermi surface reconstruction fully establishing at lower doping. Finally we show that phase fluctuations 
of the CDWs, which are enhanced in the presence of strong correlations near the Mott insulating phase,
naturally account for  the disappearance of the CDWs at low doping with yet another QCP.
\end{abstract}
\date{\today}
\maketitle

\section{Introduction}
The phase diagram of high-temperature superconducting cuprates (HTSC) is quite rich and indicates the coexistence of and
competition between different physical mechanisms. First of all strong electron-electron correlations give rise to
 an antiferromagnetic (AF) Mott insulating phase when the CuO$_2$
planes are half-filled (one hole per unit cell). Although the AF phase is rapidly disrupted by doping, nearly critical spin fluctuations 
extend their action in the metallic phase. A major distinctive feature is then a pseudogap (PG) phase occurring below the
doping-dependent temperature $T^*(p)$ (in the prototypical HTSC family YBa$_2$Cu$_3$O$_y$ (YBCO) $T^*\sim 220-250$ K at doping $p\sim 0.1-0.12$ and rapidly
decreases merging with the SC $T_c$ around optimal doping and seems to vanish at $p\approx 0.19$).  An intense debate is ongoing about the
very nature of this PG phase with  essentially two opposite points of view. Since the early times the idea was put forward (mostly by P. W. Anderson, \cite{anderson1})
 that these systems are strongly correlated doped Mott insulators, where the large electron-electron repulsion and the consequent
 short-range AF correlations, inside the low dimensional layered structure of the cuprates render these systems intrinsically different
from standard metals ruled by the Landau Fermi liquid (FL) paradigm.
In this framework many variants have been proposed ranging from the Luttinger liquid
and the Resonating-Valence Bond paradigms \cite{anderson1} and its gauge-field relatives \cite{nagaosa_lee}, to the doped d-wave Mott  state\cite{YRZ} 
to the fractionalized Fermi liquid \cite{review_sachdev}.
The occurrence of this non-FL phase may imply a drastic rearrangement
of the fermionic states: while far from the Mott state a FL is present with a large Fermi surface containing $n_h=1+p$ holes
per unit cell in the CuO$_2$ planes, approaching the Mott state the metallic character is given by just $p$ carriers residing in four hole pockets
in the so-called nodal regions $(\pi/2a)(\pm 1, \pm 1)$ of the Brillouin zone\cite{nagaosa_lee,YRZ,review_sachdev}.In this context 
the PG phase is clearly related to the ``Mottness'' of the metallic state and, noticeably, $T^*(p)$ marks a true phase transition towards
the non-FL pseudogapped phase.

The opposite point of view is that in two dimensions strong correlations and the short-range
AF correlations of doped Mott insulator are not enough {\it per se}  
to spoil the Landau FL \cite{metzner} and the anomalous behavior of the metallic cuprates should be
ascribed to the proximity to some form of instability ending at zero temperature into a second-order transition
(quantum critical point, QCP). In this case the incipient order, which at low or zero temperature has an intrinsic quantum 
(and therefore dynamic) character produces strong long ranged and long lived fluctuations. In turn these mixed quantum-thermal  fluctuations
mediate strong  scattering between the quasiparticles spoiling the FL character of
(some of) the quasiparticles, possibly mediating a strong superconducting pairing. In this ``quantum criticality'' scenario a crucial role is obviously
played by the type of order that the system would like to realize. Although many proposal have been put forward, the old evidences of charge density waves (CDW) 
\cite{reviewQCP1,reviewQCP2,kivelson_review} have been strongly revived by the recent ubiquitous observations of charge
density waves (CDWs) in all HTSC families.
\begin{figure*}[htbp]
\includegraphics[angle=-0,scale=0.65]{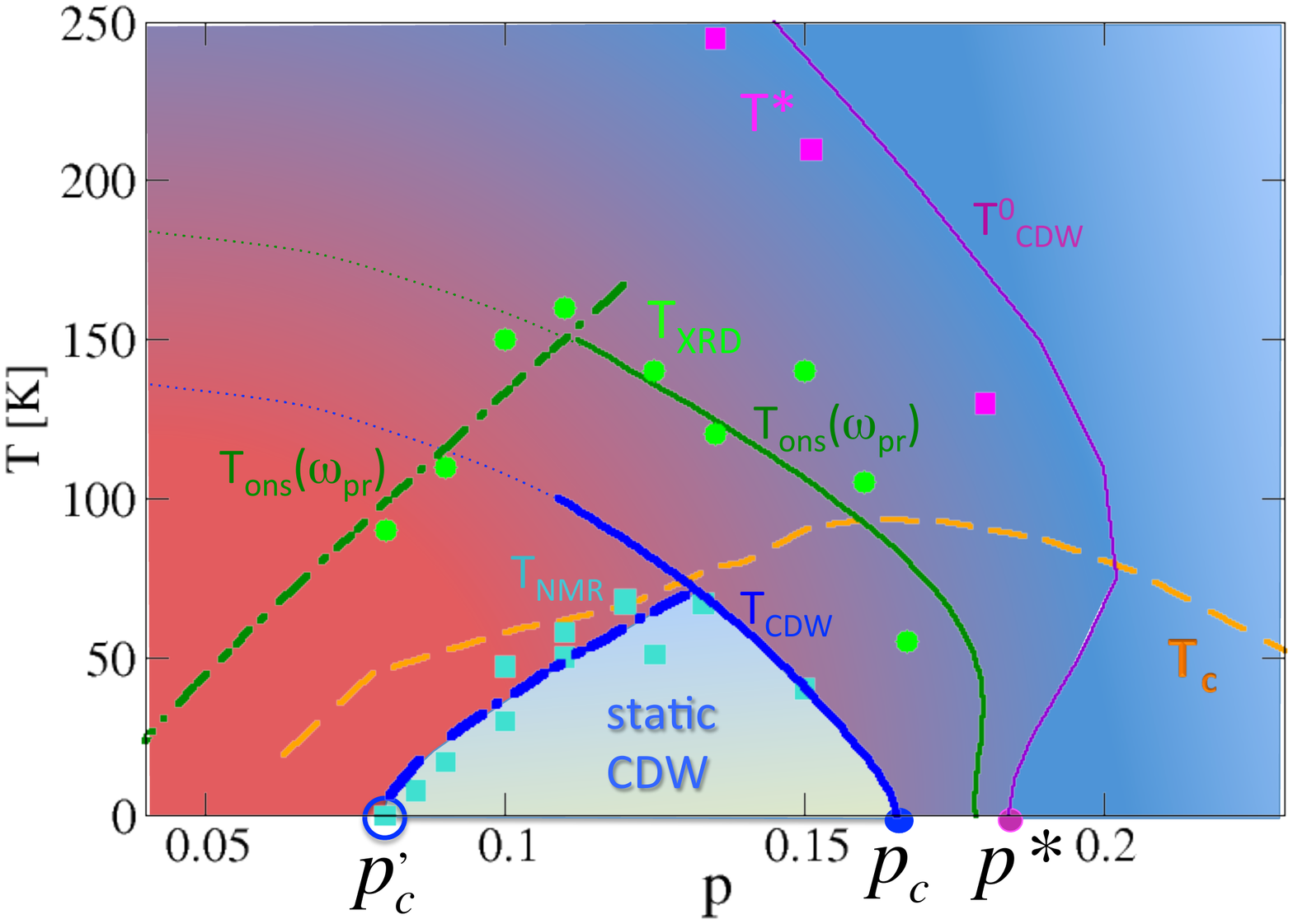}
\caption{Summary YBCO phase diagram. Increasing reddish hue corresponds to more well-defined CDW amplitude correlations.
 Experimental curves: Orange dashed line: superconducting critical temperature $T_c(p)$; Magenta squares: 
pseudogap crossover line (from Ref.\,\onlinecite{ando}, but see also Fig. 6 of Ref.\,\onlinecite{cyrchoiniere}); Green 
circles: CDW onset temperature from X-ray scattering experiments \cite{blancocanosa,hucker3}; Light blue 
squares: static CDW order as obtained from NMR and Hall experiments \cite{badoux}. On the optimal/overdoped 
side the magenta solid line tracks the mean-field CDW transition temperature identifying the region below 
which CDW fluctuations (of both modulus and phase) become prominent. The blue solid line is the static 
CDW transition line taking quantum/thermal fluctuations beyond mean-field into account. The intermediate 
green solid curve reports the dynamical CDW onset for probes having characteristic frequency 
$\omega_{pr}=50$\,K (see text). On the underdoped side the blue dot-dashed line represents the CDW 
transition ruled by phase fluctuations of the CDW order parameter (see text). In the region below the 
blue solid line (and of the thin dotted line as a guide to the eye) and above the blue dot-dashed line 
the modulus of the CDW order parameter is finite, while phase fluctuations destroy the static order 
(possibly leaving a vestigial nematic order). The thick  green dot-dashed line marks the dynamical onset 
observed by dynamical probes. This line is similar to the green solid line, but only phase fluctuations 
are involved here. The magenta dot marks the endpoint of the pseudogap crossover line at 
$p^*\approx 0.19$, identified with the endpoint of the mean-field transition, $p^*=p_c^0$. The solid 
blue dot is the true CDW-QCP at $p_c\approx 0.16 $ (with fluctuations taken into account), while the blue 
circle at $p'_c\approx 0.08$ is the low-doping QCP where the CDW phase stiffness vanishes.}
\label{phase-diag}
\end{figure*}

The observed CDWs (see, e.g., \cite{sound,taillefer,sebastian,badoux,julien1,julien3,yazdani,davis,ghiringhelli,
chang2,cominscience2014,blancocanosa,hucker3}) appear as a long-ranged phase under special 
circumstances only, like when high magnetic fields suppress superconductivity \cite{sound,julien1}.
Fig. \ref{phase-diag} reports the phase diagram of YBCO, where the light-blue 
squares mark the onset of this static CDW order as detected by NMR and Hall experiments (\cite{badoux} and references therein). 
Dynamical onset of CDWs, theoretically predicted long ago 
\cite{CDG,reviewQCP1,reviewQCP2,andergassen}, has been observed via X-ray spectroscopies 
\cite{ghiringhelli,chang2,blancocanosa} (green circles in Fig.\,\ref{phase-diag}). Recent Hall 
effect measurements indicate that a Fermi-surface reconstruction takes place in YBCO
\cite{badoux} and La$_{2-x}$Sr$_x$CuO$_4$ (LSCO) \cite{badoux-LSCO} at low temperature starting at doping 
$p'_c\approx0.08$ and ending into a QCP at $p_c\approx 0.16$. On the other hand, the number of carriers 
changes rapidly between $p_c\approx0.16$ and $p^*\approx0.19$, where $T^*(p)$ extrapolates, with a clear
connection to the above ``Mottness'' vs ``quantum-criticality'' issue discussed above. $p_c$ and $p^*$ are
distinct and from the ``Mottness'' point of view one may argue that, while $p^*$ marks the physical onset 
of a novel non-FL metallic phase, $p_c$ is the start of the CDW phase, which is a mere ``epiphenomenon''
occurring ``on top'' of the more fundamental non-FL state. Conversely, supporters of the SC point of view may
stress the close proximity of $p_c$ and $p^*$ claiming that CDW play the central role with their 
QCP at $p_c=0.16$, while $p^*$ is a crossover doping marking the region where strong CDW fluctuations 
appear before the QCP is met.

We exploit the recent experimental data to shape a coherent scenario based on the ``quantum critical'' point of view,
which rationalizes the following issues: 
a) How  the different CDW onset curves and corresponding seemingly different QCPs are related to one another? 
b) Are the CDWs related or unrelated to the pseudogap and to its onset temperature $T^*(p)$? 
c) can CDWs account for the rapid changes of the Hall number seen in experiments? 
d) Which is the mechanism leading to the disappearance of the CDWs at low doping with yet another QCP located 
at $p'_c\approx 0.08$? We concentrate our analysis on YBCO, for which the most complete set of experiments has 
been collected. Our results in comparison with experiments are summarized in the phase diagram of 
Fig.\,\ref{phase-diag}.

Our paper is organized as follows: Section II initially revisits the frustrated phase separation mechanism underlying
the formation of incommensurate CDW in strongly correlated systems. 
From this starting point a new analysis is carried out explaining why the strong correlations of HTSC favor
the occurrence of CDW along the observed directions of the Cu-O bonds. The dynamical character of the CDW fluctuations is
then analyzed to explain why probes with different characteristic timescales may detect different CDW onset temperatures. 
Section III provides a possible explanation for the
rapid change in carrier density observed by Hall experiments based on the strongly anisotropic scattering induced by CDW fluctuations.
In Section IV, again starting form the strongly correlated character of cuprates, we propose a mechanism to explain why CDW tend to 
weaken and disappear in the low-doping region of the phase diagram. In Section V  we discuss our findings and we present our concluding remarks.

\section{The optimal/overdoped phase diagram: Dynamical CDW crossovers and CDW transition}

\begin{figure}[htbp]
\includegraphics[angle=-0,scale=0.3]{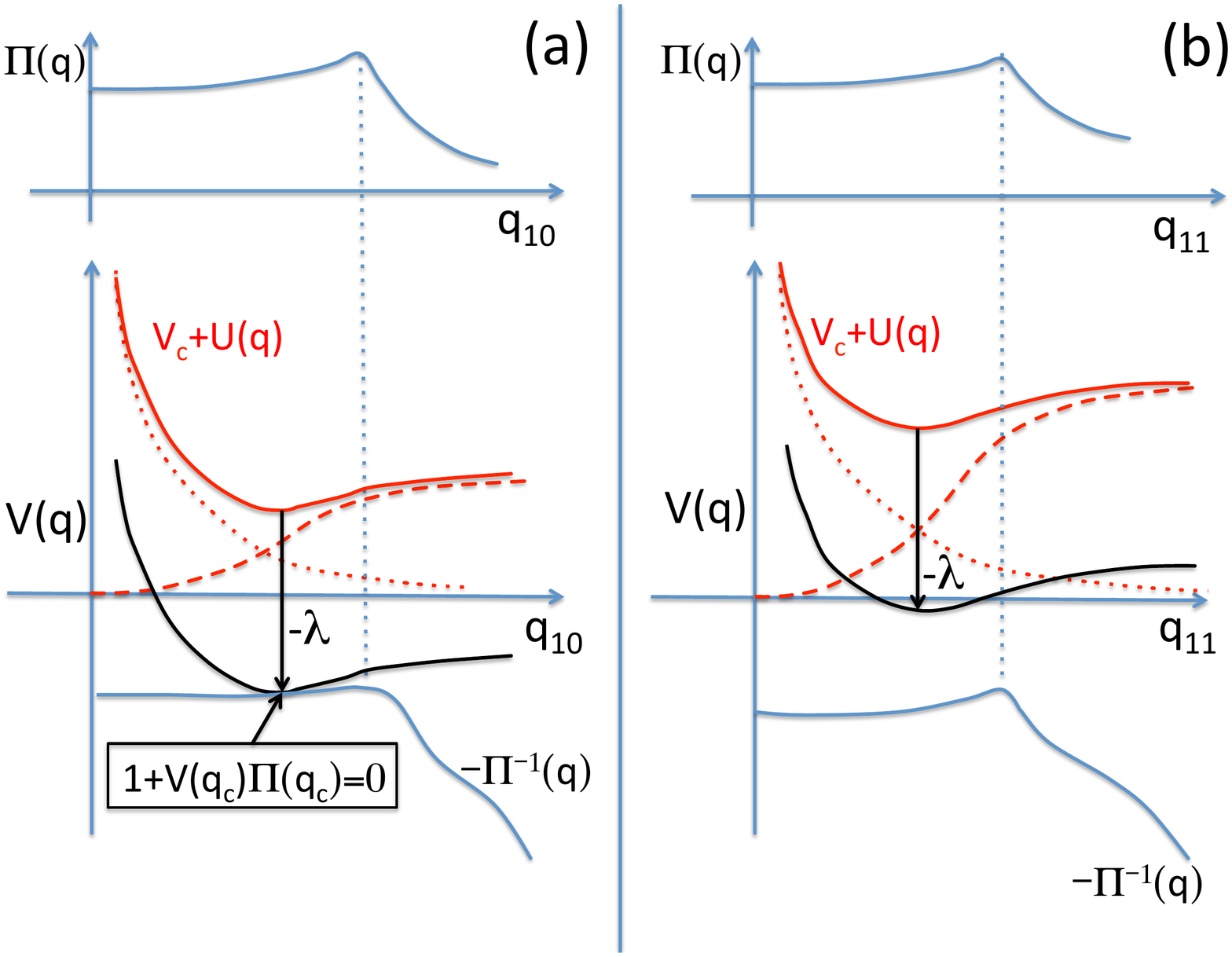}
\caption{Sketch of the {\it ingredients} determining the CDW instability and its wavevector. (a) Upper 
panel: Lindhard polarization function $\Pi(\qvec)$ with $\qvec$ along the $(1,0)$ or $(0,1)$ 
directions. Lower panel, black solid line: Residual interaction $V(\qvec)$ among quasiparticles, arising 
from the sum of the Coulomb repulsion (red dotted line), the short-range residual repulsion $U(\qvec)$ 
(red dashed line), and a nearly momentum independent attraction $\lambda$ shifting downwards the repulsive 
interaction. The instability occurs when $V(\qvec_c)=-\Pi^{-1}(\qvec_c)$. (b) Same as (a), but in the
$(1,\pm 1)$ directions, where $U(\qvec)$ is stronger and the same $\lambda$ is not enough to render the 
system unstable: $V(\qvec)>-\Pi^{-1}(\qvec)$ for all $\qvec$'s.}
\label{effint}
\end{figure}

Before addressing the above a)-d) issues, we revisit the frustrated phase separation  mechanism, which 
was proposed long ago\cite{RGCDK,EK} as the formation mechanism of CDW, giving rise to a QCP around optimal doping \cite{CDG}.
The region around optimal doping is naturally described within a Fermi-liquid picture and CDWs occur as a 
second-order instability. The correlated character of the Fermi liquid is described within a standard 
Slave-boson/Gutzwiller approach, where the residual interaction among the quasiparticles 
$V(\qvec)= U(\qvec)-\lambda+V_c(\qvec)$ arises from three distinct contributions (see Appendix A):
$U(\qvec)$ is a short-range residual repulsion stemming from the large repulsion of a one-band Hubbard model, 
$\lambda$ is a local short-range attraction triggering charge segregation, due to a local phonon \cite{CDG}, 
to the instantaneous magnetic interaction present in doped antiferromagnets \cite{EKL}, or to both mechanisms,
and $V_c(\qvec)$ is the long-range part of the Coulomb repulsion. Notice that even if phonons are involved in 
$\lambda$, they are not directly related with pairing. The screening processes are described by the Lindhard 
polarization bubble $\Pi(\qvec,\omega_n)$ for quasiparticles having a renormalized band structure fitting 
the dispersion obtained from angle-resolved photoemission spectroscopy (ARPES) experiments. The CDW instability 
is found as a divergence of the density-density correlation function, when 
$1+V(\qvec)\Pi(\qvec,\omega_n)=0$ at $\omega_n=0$ and $\qvec=\qvec_c$ 
\cite{CDG,reviewQCP1,andergassen}. For its pictorial representation see Fig.\,\ref{effint}.
In our mechanisms strong correlations favor the CDW instability along 
the (1,0) or (0,1) Cu-O bond directions, in agreement with hard X-ray experiments \cite{hucker3}.
Indeed, these are the directions along which the short-range repulsion (see Fig.\,\ref{effint}(a) and Appendix A) 
is smaller, making the instability of the frustrated phase separation due to the local effective attraction 
$\lambda$ easier. This shows that the frustrated phase separation mechanism naturally exploits the strongly correlated nature
of HTSC to account for the occurrence of CDW along the  ubiquitously observed (1,0) or (0,1) directions.

Expanding $V(\qvec)$ and $\Pi(\qvec,\omega_n)$ around $\qvec=\qvec_c$ and $\omega_n=0$ one obtains the 
standard quantum-critical charge-fluctuation propagator (see Appendix A)
\beq
D(\qvec,\omega_n)=\left[m_0+\nu(\qvec)  +|\omega_n| +\omega_n^2/\overline{\Omega} \right]^{-1},
\label{fluctuator}
\eeq
where $m_0\propto \xi_0^{-2}\propto 1+V(\qvec_c)\Pi(\qvec_c,0)$ is the mean-field {\it mass} of the 
fluctuations, $\xi_0$ is the mean-field CDW correlation length, $\nu(\qvec)\approx \bar\nu |\qvec -\qvec_c|^2$, 
$\bar\nu$ is an electronic energy scale (we work with dimensionless momenta, measured in inverse lattice 
spacings $1/a$), and $\overline{\Omega}$ is a frequency cutoff. The mean-field instability line $T^0_{CDW}(p)$ 
(magenta solid line in Fig.\,\ref{phase-diag}) is characterized by a vanishing $m_0$. This is the well-known 
frustrated-phase-separation instability \cite{EK,RGCDK} underlying the formation of CDWs near optimal 
doping \cite{CDG,reviewQCP1,andergassen}. We notice that, 
although $\lambda$ might have a magnetic contribution, the present mechanism of CDW formation, contrary to 
other proposals \cite{kontani,sachdev,efetov,chubukov}, does not require the proximity to a magnetic QCP. The same 
CDW-mediated interactions are also active in the Cooper channel, providing a high-temperature 
$d$-wave pairing mechanism \cite{perali}. Therefore, this region of the phase diagram is characterized 
also by the gradual onset of CDW-mediated pairing fluctuations, with reducing temperature and/or doping.

The microscopic parameters of the model are adjusted to fix the mean-field QCP $p_c^0\approx 0.19$, while 
the temperature dependence arises from the Lindhard function, {\it without further adjustable 
parameters}, and tracks the pseudogap onset line $T^*(p)$. In our scheme, below this line CDW long-range 
order would occur, were it not for quantum-critical fluctuations especially due to the quasi-bidimensionality of the cuprates. 
In real systems, where strong fluctuations 
are present, below this line CDW fluctuations become increasingly long-ranged and 
long-lived, and start to suppress the quasiparticle states preferably in the antinodal regions
of the Fermi surface, contributing 
to a gradual formation of the Fermi arcs, likely due to the interplay of CDW and CDW-mediated pairing fluctuations
(see, e.g., Ref.\,\onlinecite{allais}). 
This explains the connection of our theoretical mean-field line $T^0_{CDW}(p)$ with $T^*(p)$ [issue b) in the Introduction]. 

The fluctuation suppression of the mean-field critical line $T^0_{CDW}(p)$ is obtained by the self-consistent 
evaluation of the correction to the mean-field mass $m_0$, due to the fluctuator Eq.\,(\ref{fluctuator}) 
(see Methods, Fig. \ref{fig-feynman}),
\beq
m=m_0+uT\sum_n\int_0^\Lambda \mathrm d\nu \,\frac{N(\nu)}{m+\nu+ |\omega_n|+\omega_n^2/\overline{\Omega}},
\label{renmass}
\eeq
where $u$ is the strength of the coupling between CDW fluctuations, $N(\nu)$ is the density of states 
corresponding to the dispersion law $\nu(\qvec)$ and $\Lambda$ is an ultraviolet cutoff, corresponding to a 
momentum cutoff $\bar{q} \sim 1/a$. 

One can numerically solve the self-consistent expression (\ref{renmass}), finding the conditions for the 
instability $m=0$ and the dependence of $m$ on temperature and doping. In a 
two-dimensional system, as a single CuO$_2$ plane of cuprates, $N(\nu)$ is constant for $\nu\to 0$. This 
leads to a finite shift of the two-dimensional QCP at $T=0$, but the correction to the critical line at 
finite $T$ is divergent, consistently with the absence of long-range order in two 
dimensions for a two-component order parameter (as in the incommensurate CDW case). 
However, cuprates are layered systems and the planes are weakly coupled. This allows for an ordering 
at finite temperature $T_{CDW}(p)$, which arises in the solution of Eq.\,(\ref{renmass}) 
for $m=0$, provided $N(\nu)$  assumes a three-dimensional form $N(\nu) \sim \sqrt{\nu}$ below some energy 
scale $\nu_\perp$ related to the inter-plane coupling (see Methods). $T_{CDW}(p)$ is so reduced (see 
Fig.\,\ref{phase-diag}) with respect to the mean-field line $T_{CDW}^0(p)$ that it occurs below the 
superconducting dome. Superconductivity therefore appears as the stabilizing phase against CDW long-range 
order. This explains why the experimental data corresponding to long-range 
CDWs are only detected for magnetic fields large enough to weaken the superconducting phase \cite{sound}. 
These experiments also allow to estimate $\nu_\perp$. Fig.\,\ref{phase-diag} displays the $m(T_{CDW})=0$ 
blue line obtained with $\nu_\perp=10$\,K. 

We now address the issue a) of why different probes identify different CDW onset temperatures. The 
key point is the dynamical character of the CDW fluctuations. A probe with long characteristic timescale 
(like, e.g., NMR or NQR) will only detect static order, otherwise the fluctuating CDWs average to zero during 
the probing time. This is why these probes identify a true phase-transition line $m(T_{CDW},p)=0$ at high 
magnetic field (of course, if in real systems pinning intervenes to create locally a static order, this can 
be detected by local static probes even at larger temperatures and low magnetic fields. 
This seems to be the case in recent NMR experiments \cite{julien3}). On the other 
hand, a fast probe with a short probing time $\tau_{pr}$ takes a fast snapshot of the fluctuating system
and finds a seemingly higher transition temperature when the CDW order is still dynamical, as long as the CDW 
characteristic timescale $\tau_{CDW}\propto \xi^2\propto m^{-1}$ is longer than $\tau_{pr}$, thus acting as an infrared 
cut-off and diminishing the reduction effect of the fluctuations on the the onset temperature. We identify 
the dynamical onset line as the line where $m\approx\omega_{pr}=\tau_{pr}^{-1}$. 
The green solid line in Fig.\,\ref{phase-diag}, represents the dynamical onset of CDWs observed 
with X-ray spectroscopy with a mass $m\approx\omega_{pr}=50$\,K, thus solving the experimental puzzle a).

\begin{figure}[htbp]
\includegraphics[angle=-0,scale=0.35]{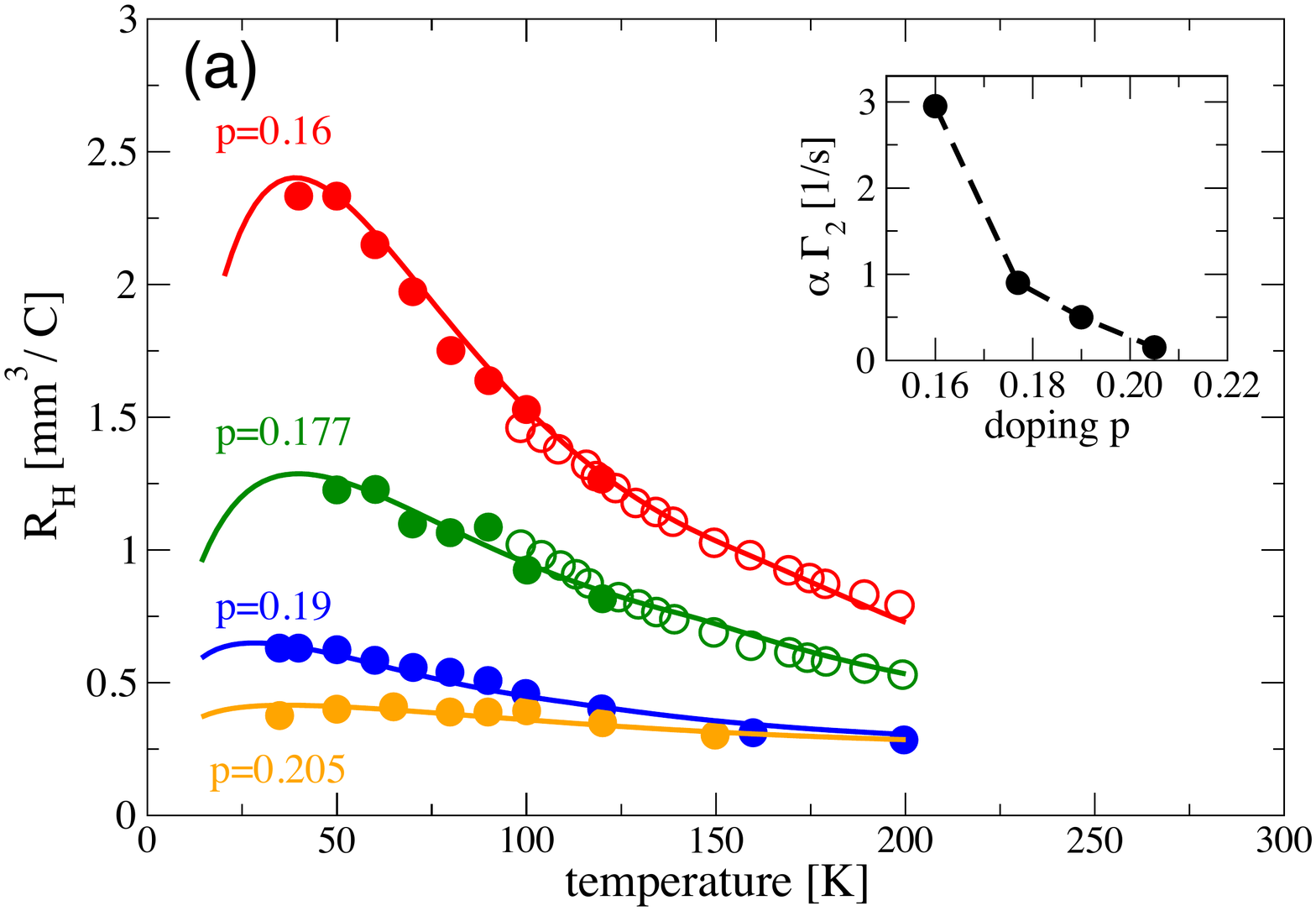}\\
\includegraphics[angle=-0,scale=0.35]{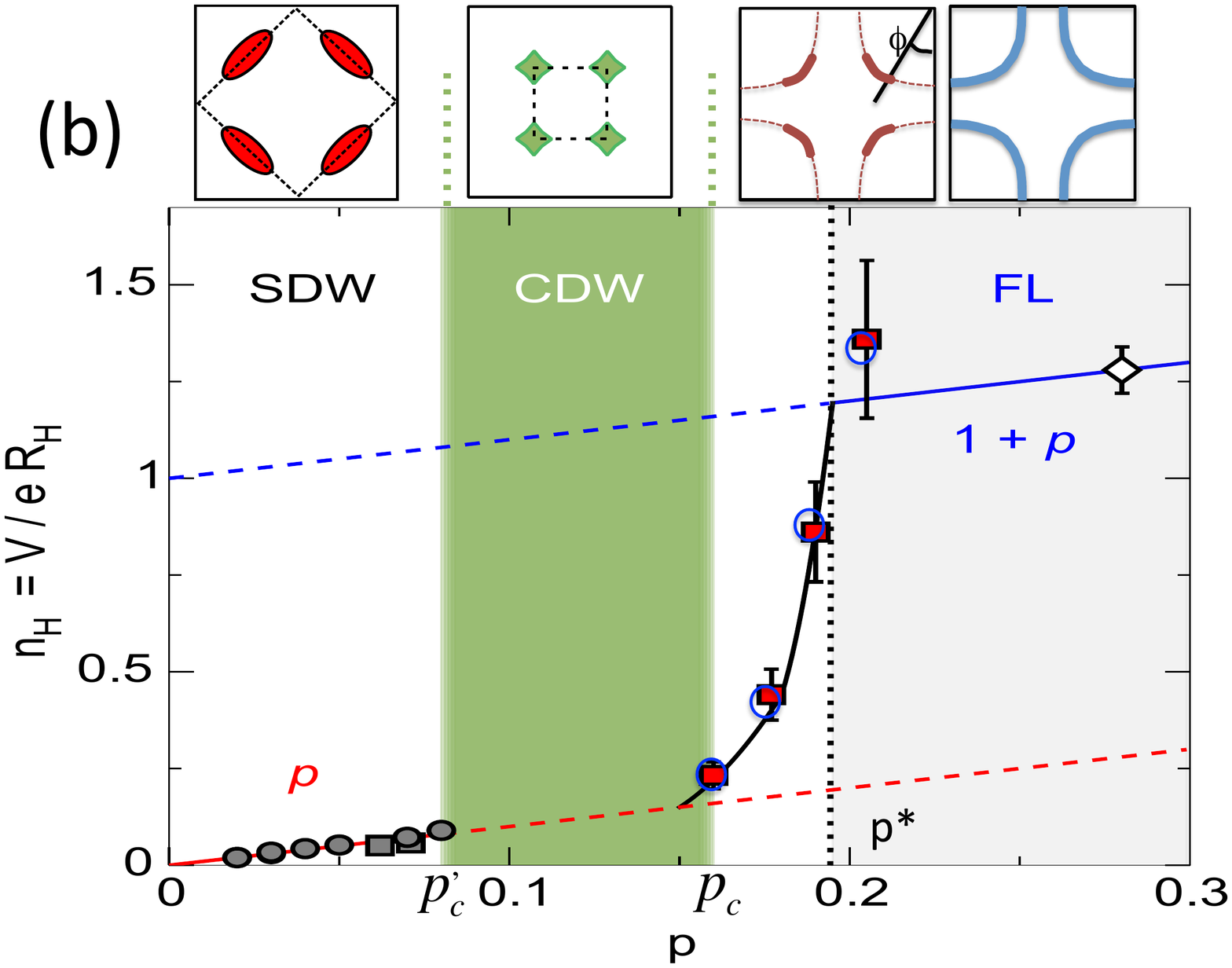}
\caption{Analysis of Hall constant and hole density within the CDW-QCP scenario.
(a) Fits of $R_H$ vs. $T$ of the low-field (empty circles) and high-field (filled circles) measures at 
various dopings (data taken from Ref.\,\onlinecite{badoux}) using the anisotropic scattering model of 
Eqs.\,(\ref{gamma1}) and (\ref{gamma2}). The anisotropy parameter $\alpha \Gamma_2$ is reported in the 
inset. (b) (Adapted from Ref.\,\onlinecite{badoux}) Doping evolution of the hole density extracted from 
$R_H$ measurements in Ref.\,\onlinecite{badoux}. The experimental points in the range $0.16\le p \le 0.205$ are 
the red rectangles with error bars, while the blue circles are the values of $n_H$ extracted from the fits 
based on the microscopic anisotropic scattering model. The Fermi surfaces in the various doping regimes are 
schematically depicted above: (from left to right) hole pockets in the strongly correlated region near the 
Mott transition, Fermi surface reconstruction with electron pockets in the CDW region; large Fermi surface 
for $p>p_c\approx 0.16$ with dashed (hot) regions where for $T<T^0_{cdw}$ the anisotropic CDW scattering
suppresses the quasiparticle states; for $T<T_{pair}$ this suppression is enhanced by CDW-mediated pairing fluctuations
leading to the formation of Fermi arc features; large Fermi surface in the region far from the CDW-QCP, where 
CDW scattering is weak and the Fermi-surface states are cold.
}
\label{Hall}
\end{figure}

\section{ CDW-QCP interpretation of Hall transport} 

According to issue c), recent Hall experiments \cite{badoux}
show a rapid increase of the Hall number in a narrow doping range
from $n_H=p$ at $p\lesssim 0.16$ up to $n_H=1+p$ for doping
larger than the pseudogap zero temperature onset point $p^*$. Such crossover has been
attributed to a large Fermi surface at $p>p^*$
as observed with ARPES and the formation of hole pockets for $p<p^*$
due to either the establishment of a new metallic phase like a d-wave Mott insulator \cite{YRZ,storey} or a 'topological metal' \cite{sachdev2}, or some
more conventional kind of order as e.g. a spin spiral \cite{eberlein}.
In all these scenarios $p^*$ is completely {\it unrelated} to CDWs
whose QCP is then placed at the lower doping value $p_c$.

Here, instead, we keep a minimal framework
showing that the crossover behavior of the Hall number
can be explained by the increasingly strong fluctuations starting below $T^*(p)$ and
approaching the nearby CDW-QCP.  In particular, we exploit the strong 
momentum dependence of the effective quasiparticle interaction mediated by the CDW propagator 
Eq.\,(\ref{fluctuator}) which naturally splits the Fermi surface in {\it hot} and {\it cold} regions: The 
quasiparticles in the hot regions ($\phi=0,\pi/2$) are connected by $\qvec \sim \qvec_c$ and 
interact strongly, while the scattering in the cold regions stays weak.
This results in a marked anisotropy of the quasiparticle 
scattering rate  \cite{sulpizi}
\begin{equation}
\Gamma(\phi)=\{\Gamma_{max}^{-1}+[\Gamma_0 + \Gamma_{\Sigma}(\phi)]^{-1}\}^{-1},     
\label{gamma1} 
\end{equation}
with
\begin{equation}
\Gamma_{\Sigma}(\phi)= \mathrm{Im} \frac{2\, T\, \Gamma_2(\phi)}{1+\sqrt{1-i\frac{T}{M(\phi)}}},
\label{gamma2} 
\end{equation}
and $\Gamma_2 (\phi)=\Gamma_2 \lbrack 1+\alpha\cos^2(2\phi)\rbrack$ .
The expression of $\Gamma_{\Sigma}(\phi)$ is derived from the quasiparticle self-energy near hot spots in 
models with spin \cite{abanov} or charge \cite{sulpizi} nearly critical fluctuations. Here,
$M(\phi)=m(T)+\bar{\nu} \sin^2(2\phi)$ is the energy scale below which the quasiparticles have a Fermi-liquid 
behavior and is minimal at the hot spots. The anisotropy of $\Gamma(\phi)$ is enhanced by approaching the 
CDW criticality, where $m(T)$ vanishes. 
The resulting fits of $R_H$ as a function of temperature are reported 
in Fig.\,\ref{Hall}(a), while the related hole densities are reported in Fig.\,\ref{Hall}(b) by the blue 
circles, in good agreement with the experimental values (red squares) \cite{badoux}. Interestingly, we also 
find a good agreement with the resistivity curves [see Fig.\,\ref{figMethHall2} in Methods] in contrast to a similar analysis in Ref. \cite{badoux}.
The inset of 
Fig.\,\ref{Hall}(a) also shows that the anisotropic component of scattering strongly increases 
with approaching $p_c\approx 0.16$ so that the dominating contribution
to $n_H$ is coming from the nodal regions.

 The anomalous singular scattering is a general consequence of the CDW-QCP and it {\it must} occur when this is approached, 
 as it has indeed been measured by quantum oscillations in \cite{ramshaw}, 
and corresponds to the region of Fermi surface reconstruction and of anomalous behavior of the Hall number.
Hence, coming from high temperatures our CDW-QCP scenario predicts a significant increase of $R_H$ 
upon crossing the pseudogap critical temperature
when the CDW scattering starts to
deplete spectral weight from the antinodal Fermi surface states \cite{epj2000} 
$T^0_{cdw}$ in agreement with ARPES experiments \cite{kaminski}. 
However, it is important to note that our analysis does not incorporate pairing fluctuations which 
have been shown \cite{kaminski} to be relevant in a certain
temperature window  $T_c<T<T_{pair}< T^0_{cdw}$ and which would further
increase the anisotropic scattering via the formation of antinodal gaps in
the electronic spectrum (the relevant interplay
of CDW and pair fluctuations has already been noticed, e.g. in Ref. \cite{allais}).
In particular, if long-range superconducting
order is suppressed by strong magnetic fields these fluctuations
extend down to $T=0$ in contrast to the
anisotropic scattering from Eq. (\ref{gamma2}) which vanishes in this limit (even though the vanishing of 
$m(T)$ tempers this reduction).
Therefore, upon lowering temperature the increasing contribution of pairing fluctuations
will counteract the decrease in the CDW anisotropic quasiparticle
scattering and thus eliminate the downturn of the $R_H$ curves in Fig. \ref{Hall} in the
limit $T\rightarrow 0$ as is observed in Nd-codoped LSCO \cite{collignon} where $R_H$
can be measured to much lower temperatures than in YBCO where the measures end at $T=40$ K.
\vspace{1 truecm}

\section{The underdoped phase diagram: the role of CDW phase fluctuations}
\begin{figure}[htbp]
\includegraphics[angle=-0,scale=0.32]{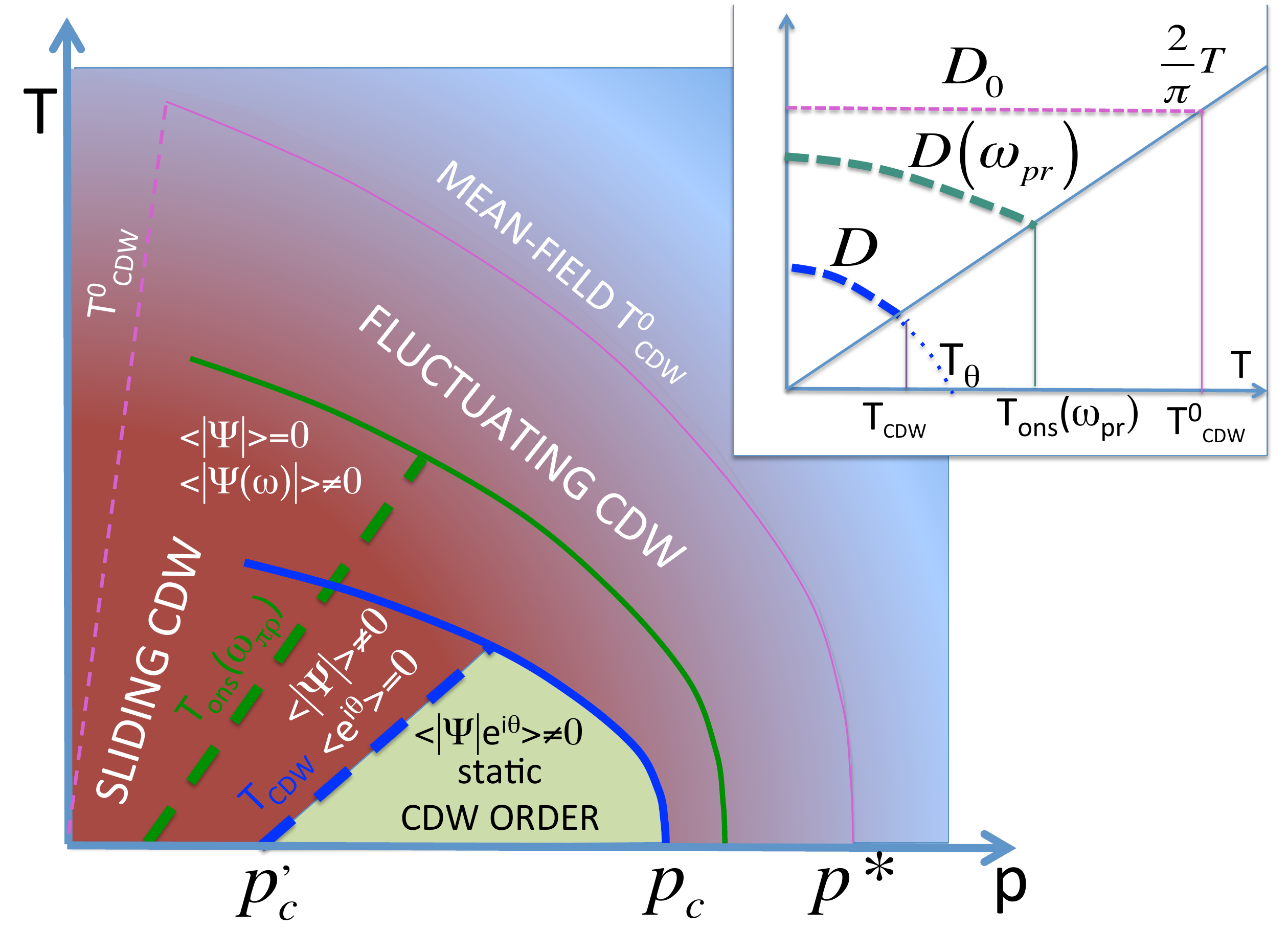}
\caption{Schematic view of the phase diagram. The solid magenta line on the right is the 
mean-field instability line, while the dashed one on the left is the BKT transition temperature as obtained 
from the {\it bare} CDW phase stiffness $T_{CDW}^0=\frac{\pi}{2}D_0$ (see magenta dashed line and thin solid 
blue line in the inset). The thick solid blue line is the transition line obtained by 
fluctuations correcting the mean-field instability line. Below this line the amplitude of the CDW order 
parameter is finite $\langle \vert \Psi (0)\vert \rangle\ne 0 $. The analogous line on the left is dashed 
and arises from the thermal and quantum phase fluctuations reducing the phase stiffness (see blue dashed 
line in the inset at fixed doping). The intermediate lines arise when fluctuation corrections are cut off 
in the infrared limit. The green solid line is for amplitude and phase fluctuations, partially reducing 
$T_{CDW}^0$.  Below this line the {\it static} average of the CDW order parameter amplitude vanishes, but 
it is finite for long time intervals tested by fast probes, 
$\langle \vert \Psi (\omega>\omega_{pr})\vert \rangle\ne 0$: The green dashed line is the 
onset temperature $T_{ons}$ corresponding to a phase stiffness, which is less reduced because of an 
infrared cutoff to phase fluctuations (also reported as a green line in the inset for a fixed doping). 
Inset: CDW phase stiffness at fixed doping as a function of temperature. The magenta dashed line is the 
bare stiffness $D_0$. The blue dashed is the the stiffness $D$ corrected by phase fluctuations. 
Disregarding the discontinuity at the BKT transition, the phase stiffness vanishes at $T_\theta$ (dotted 
blue line). Green dashed line: same as the blue line, but the fluctuations have
an infrared momentum cutoff $\omega_{pr}/c_\theta$ to mimic the stiffness on timescales shorter than 
$\omega_{pr}^{-1}$.
}
\label{scheme}
\end{figure}

In all the models proposed so far (including ours), were it not for the competing superconducting phase, 
the critical line for static CDWs monotonically increases with underdoping and saturates to a finite value 
at low doping (see Fig.\,\ref{phase-diag} and Fig.\,\ref{scheme} below the solid blue line). This is at
odds with the experiments, where CDWs appear below a dome-shaped critical line ending into a QCP at low doping 
$p_c'=0.08$  \cite{sebastian2010}. In our model we solve this inconsistency taking into account the role of 
the dynamical fluctuations of the phase $\theta$ of the CDW order parameter 
$\Psi(\rvec)=\vert \Psi (\rvec) \vert e^{i \theta(\rvec)}$ at low doping.

A well-known feature of CDWs \cite{gruener} is that their phase stiffness is 
proportional to the strength of the metallic character (like, e.g., high Fermi velocity, high
quasiparticle DOS, etc.), which is strongly reduced in the proximity of the 
Mott transition. 
Thus the experimentally observed 
reduction of the CDW critical (or onset) temperatures in the underdoped region appears as a crossover
from a transition ruled by the vanishing of the amplitude of the CDW order parameter 
$\langle \Psi \rangle=\langle \vert \Psi (\rvec) \vert \rangle =0$
to a transition controlled by the suppression of the stiffness, due to
fluctuations of the phase $\theta$. In this region (below the solid blue line and above the dashed 
blue line in Fig.\,\ref{scheme}) the translational symmetry is restored by the sliding motion of the CDW, 
with the sliding CDW possibly oriented along the a or b direction of the 
CuO$_2$ planes, thereby marking a nematic breaking of the $C_4$ symmetry of the lattice. This situation 
is reminiscent of the so-called {\it vestigial} charge order \cite{kivelson-PNAS,capati2015}, obtained in a 
different context. 

The effect of the phase fluctuations on the CDW dynamics is customarily described by the $XY$-like action
\[
\mathcal{S}_{XY}=\frac{1}{8}\sum_{\qvec,\omega_m}\left[D_0 q^2+\chi\omega_m^2\right] \theta(\qvec,\omega_m) 
\theta(-\qvec,\omega_m),
\]
where $D_0$ is the bare stiffness, and the coefficient $\chi$ determines the speed $c_\theta=\sqrt{D_0/\chi}$  
of the phase fluctuations \cite{gruener}. As long as the CDW critical temperature is larger 
than the interplane coupling ($\approx 10$\,K in the optimal doping region), we can take 
this transition to be of the Berezinski-Kosterlitz-Thouless (BKT) type. According to the above arguments, 
the phase stiffness $D_0\propto p$ and therefore the bare critical temperature of the BKT 
transition obtained from the usual condition $T^0_{CDW}=\frac{\pi}{2}D_0$ (see magenta line in the inset 
of Fig.\,\ref{scheme}) is also proportional to doping $T^0_{CDW}=\bar{T}\, p$ (magenta dashed line in 
Fig.\,\ref{scheme}).

However, similarly to what happens on the high-doping side to the mean-field critical temperature, whatever 
is the bare $T^0_{CDW}$, this is reduced by the quantum and thermal phase fluctuations. Within the XY model 
with sound-like phase fluctuation modes one obtains the perturbative 
expression of the depleted stiffness \cite{benfatto} 
\beq
D=D_0-\frac{u_\theta \, c_\theta}{2\pi \Lambda_\theta^2} \int^{\Lambda_\theta}_0 \mathrm d q\, q^2 
\left[1+b(c_\theta q)\right]
\label{phase_fluct}
\eeq
where $\Lambda_\theta \sim 1/a$ is a momentum cutoff (hereafter we take $\Lambda_\theta=1$) 
and $b(z)=[\mathrm e^{z/T}-1]^{-1}$ is the Bose function. Starting from different $D_0$ at the different 
doping and calculating the perturbative corrections with a doping-independent coupling $u_\theta$, we reduce 
the bare BKT transition temperature to $T_{CDW}=\frac{\pi}{2}D$, which vanishes at a value of $p=p'_c$,
thereby answering issue d) in the Introduction. The various physically relevant
combinations of the microscopic parameters can be reduced to two effective parameters only (see Appendix C), 
the position of the QCP, which can be fixed by experiments at  $p'_c\approx 0.08$ and the
slop e of the $T_{CDW}(p)$ curve. Numerically solving Eq.\,(\ref{phase_fluct}) we obtain the dashed blue line 
in Fig.\,\ref{phase-diag}.

Above the static $T_{CDW}$ the phase fluctuates dynamically and, 
similarly to what happened in the high-doping region, a fast probe may detect a 
seemingly ordered state. This occurs as long as the phase correlation time 
$\tau_\theta$ is longer than the characteristic timescale of the probe. Again, we mimic this effect 
with an infrared cutoff $\omega_{pr}$, i.e., a lower limit $\omega_{pr}/c_\theta>0$
in the integral in Eq.\,(\ref{phase_fluct}). The reduced effect of quantum and thermal fluctuations 
leads to a larger stiffness and, in turn, a CDW onset at a higher temperature. For
$\omega_{pr}=50$\,K we find the green dot-dahsed line in Fig. \ref{phase-diag}, that
matches with the same $\omega_{pr}$ the analogous line found in the optimal/overdoped region. 
The joint onset line then accounts for the dome-shaped $T_{CDW}(p)$ onset temperature found by X-ray 
scattering [issue a) in the Introduction]. 
 
The prediction of a low-doping CDW-QCP due to the vanishing of the CDW phase
stiffness rises the obvious issue of a possible direct observation of
the associated strong phase modes in the vicinity of the transition.
Since the phasons are optically active
\cite{lee-rice-anderson}, optical conductivity is a natural tool to this
purpose, but 
the contribution to the planar optical conductivity from overdamped CDW modes
has already been theoretically investigated
\cite{enss} and it turns out that this contribution is likely overshadowed by
the wealth of planar single particle excitations. 
On the other hand, the optical conductivity along the
c-axis $\sigma_c(\omega)$ could be much more suited to identify the effects.
In fact, the observed resonance in bilayer (and threelayer) cuprates \cite{bernhard} which has 
been previously attributed to transverse (Josephson)
plasma modes (TPM) \cite{vandermarel} might also be associated with
intralayer-coupled phason modes which generate a dipole moment
along the c-direction \cite{grilli16}. The expected specific doping and
temperature scaling properties \cite{enss} could be a useful
test on the origin of this optical feature.

\section{Discussion and Conclusions}

We addressed several key issues of cuprates, and yet other issues remain open. First of all, the simplest 
single-band Hubbard model neglects the intra-cell CuO$_2$ structure and therefore it makes it hard to 
address the interesting question of the internal $d$-wave structure of the CDW order parameter recently 
detected both in resonant X-ray scattering \cite{comin2015} and STM \cite{davis,fujita} experiments. 

We also ignored the effect of disorder, which may locally render the CDW fluctuations static 
\cite{vojta2009} and therefore observable by NMR well above the true static transition line \cite{julien3}.
Nevertheless our scenario is a good starting point to address these remaining open issues since it provides 
a unified scheme based on a robust and generic  frustrated phase-separation mechanism to explain
all the a)-d) issues listed in the introduction.  The direction of the critical CDW wavevector, which is a natural 
test on the validity of the various proposed CDW mechanisms, is determined here by the generic 
strong-correlation effect encoded in the residual interaction among the Fermi-liquid quasiparticles 
$V(\qvec)$. At and above optimum doping the transition is ruled by quantum and thermal CDW 
fluctuations, whereas at low doping phase fluctuations control the reduction of the stiffness and 
determine the transition. Our main result is that only one {\it bona fide} dome-shaped CDW transition 
line is present, ending into two CDW-QCPs. The low-doping one at $p'_c\approx 0.08$ marks the onset of 
a static CDW order as revealed by Fermi surface reconstruction with electron pockets. The CDW phase ends 
at higher doping $p_c\approx 0.16$, above which fluctuations destroy again the static order. The whole 
region below $T^*$ (theoretically associated with the mean-field CDW instability line) is characterized by 
increasingly stronger fluctuations for doping $p_c<p<p^*\approx 0.19$. Nonetheless, depending on the 
probe characteristic timescale, fast probes, like X-rays, can detect different onset lines $T_CDW(p)$, 
well accounted for within our dynamical CDW fluctuation approach. Extrapolating 
these lines to low temperature may erroneously lead to a misplacement of the real CDW-QCPs. This 
unifying picture is also supported by the fact that, although different experiments mark distinct onsets 
for the static CDW transition (under strong magnetic fields) and for the dynamical regime detected by X-rays, 
the same in-plane modulation vector is observed \cite{gerber} denoting their common origin. 
The experiment on the c-axis optical conductivity 
on multilayer systems, suggested in Section IV, could further support the relevance of CDW phase 
fluctuations in determining the low-doping transition line.

Within our scenario the effective quasiparticle interactions are strongly anisotropic as a result of the 
anisotropy of CDW fluctuations mediating them. Starting below $T^*(p)$, this scattering accounts for a progressive 
increase of $R_H$ as observed in Hall experiments. Although in YBCO the situation is not yet settled, this anisotropic 
scattering may be supported by other mechanisms (like pair fluctuations) to reconstruct the Fermi surface giving rise 
to arcs and pockets with effective hole number eventually tending to $p$
as it likely occurs in other cuprate families \cite{collignon}.
The interplay between CDW fluctuations and pairing is an old issue \cite{perali}, which might also 
explain why under the action of increasingly stronger 
magnetic fields the superconducting dome splits in two smaller domes (see Fig. 4 b in 
Ref.\,\onlinecite{grissonnanche}), showing that superconductivity survives longer in the critical regions 
associated to the two QCPs at $p'_c=0.08$ and $p_c =0.16$. This tight relationship between 
superconductivity and CDW criticality as it clearly emerges from this experiments further supports the 
central character of CDW quantum criticality in the cuprates phase diagram.

\vspace {1truecm}

{\bf Acknowledgements}

We acknowledge stimulating discussions with C. Castellani, K. Efetov, J. Lesueur, and J. Lorenzana. We 
acknowledge financial support from the Sapienza University Project n. C26A115HTN.

\vspace {1truecm}

\vfill \newpage
\appendix
\section{The microscopic model: mean-field  and fluctuation corrected CDW instability}

We first obtain a mean-field phase diagram within the Random-Phase Approximation (RPA) for correlated single-band 
quasiparticles. Specifically, we assume that a strong Hubbard-like local repulsion $U$ can be treated with a standard 
slave-boson/Gutzwiller (SB/G) approach \cite{arrigoni,epj2000}, leading to a Fermi-liquid scheme where the quasiparticles 
are effectively described by the Hamiltonian 
\[
H=\sum_{\kvec,\sigma} E_\kvec c^\dagger_{\kvec,\sigma} c_{\kvec,\sigma}+\frac{1}{2}\sum_{\qvec} 
V(\qvec)\rho_\qvec \rho_{-\qvec},
\]
where $c^\dagger_{\kvec,\sigma}\,(c^{\phantom{\dagger}}_{\kvec,\sigma})$ creates (annihilates) a 
quasiparticle with momentum $\kvec$ and spin projection $\sigma$. The band takes a tight-binding form 
$E_\kvec =-2t\left(\cos k_x+\cos k_y \right) +4t' \cos k_x \cos k_y-\mu$
(we take a unit lattice spacing $a=1$ on the CuO$_2$ planes of the cuprates, so 
that momenta are dimensionless), with nearest ($t$) and next-to-nearest ($t'$) neighbor hopping terms
and $\mu$ is the chemical potential.
$\rho_\qvec=\sum_{\kvec,\sigma} c^\dagger_{\kvec+\qvec,\sigma} c_{\kvec,\sigma}$ is the Fourier transform 
of the density operator, and $V(\qvec)$ is the residual density-density interaction between the 
quasiparticles. In particular, we find that 
$V(\qvec)={U}(\qvec)-\lambda + V_c(\qvec)$ arises from three contributions, each with a clear physical meaning:
\begin{eqnarray*}
{U}(\qvec)&=& {U}_0+ {U}_1\left(2-\cos q_x-\cos q_y\right) \nonumber\\
&+& {U}_2\left(1- \cos q_x \cos q_y\right) \nonumber
\end{eqnarray*}
is a short-range residual repulsion between the quasiparticles; $\lambda$ is a generic local attraction 
that would drive a phase separation instability of the quasiparticles, were it not for the presence of a 
long-range Coulomb repulsion 
\[
V_c(\qvec)=\frac{v_c}{\sqrt{G^2_\qvec-1}},
\]
which is the Fourier transform of the long-range Coulomb interaction, projected onto a single CuO$_2$ plane, 
preventing phase separation of the charged electrons. The coupling constant is 
$v_c=e^2 d/(2\epsilon_\perp)$, and
\[
G_\qvec=1+\frac{\epsilon_\parallel d^2}{\epsilon_\perp}\left(2-\cos q_x-\cos q_y\right).
\]
Here, $e$ is the electron charge, $d$ is the interlayer distance (in units of the lattice spacing on the 
CuO$_2$ planes), $\epsilon_\parallel$ and $\epsilon_\perp$ are the components of the dielectric tensor 
for a system with tetragonal symmetry, along the principal axes parallel and perpendicular to the CuO$_2$ 
planes, respectively. For the sake of definiteness we fix 
for YBCO, $\epsilon_\perp=5$, $\epsilon_\parallel=20$. As we shall see, the Coulomb repulsion 
changes the electronic phase-separation instability at $\qvec=0$ into a CDW instability at some finite 
vavevector $\qvec=\qvec_c$.

Within a SB/G approach, the quasiparticle dispersion is suppressed by the hole doping $p$, i.e.,
$t=t_{\mathrm{bare}}\,p$, $t'=t'_{\mathrm{bare}}\,p$. For YBCO, suitable values are 
$t_{\mathrm{bare}}=0.3$\,eV, $t'/t=-0.45$.

At a given doping $p$, the chemical potential $\mu$ is fixed by the equation
\[
\frac{2}{N}\sum_\kvec f(E_\kvec)=1-p,
\] 
where $N$ is the number of $\kvec$ vectors within the first Brillouin zone of the CuO$_2$ planes, and 
$f(z)=(1+\mathrm e^{z/T})^{-1}$ is the Fermi distribution function at a temperature $T$. The parameters 
of the short-range repulsion are then found as \cite{epj2000} $ U_0=-4\mu/p$,
\[
 U_1=\frac{t_{\mathrm{bare}}}{pN}\sum_\kvec \left(\cos k_x+\cos k_y\right)f(E_\kvec),
\]
and
\[
 U_2=\frac{4t'_{\mathrm{bare}}}{pN}\sum_\kvec \cos k_x\cos k_y\,f(E_\kvec).
\]

The Coulomb interaction prevents the segregation of charged quasiparticles on large scales, driven by 
$\lambda$, while leading to a finite-wavelength instability at 
$\qvec=\qvec_c$. These are the basic ingredients of the so-called frustrated-phase-separation mechanism 
\cite{EK,RGCDK} underlying the formation of CDW near optimal doping \cite{CDG}. We notice that, while 
this model has been considered in the past to describe a phonon-mediated short-range attraction, 
$\lambda$ can well describe any short-ranged (i.e., weakly momentum dependent) attraction possibly arising 
from nearest-neighbor magnetic \cite{CGK}, Coulombic \cite{GRCDK,becca1998} or/and phononic \cite{CDG,becca} 
mechanisms. The intra-band screening processes can be described by the standard quasiparticle Lindhard 
polarization bubble
\beq
\Pi(\qvec,\omega_n) =-\sum_{\kvec,\sigma}\frac{f(E_{\kvec+\qvec})-f(E_{\kvec})}{E_{\kvec+\qvec}-E_\kvec+\omega_n},
\label{bubble}
\eeq
where $\omega_n$ are bosonic Masubara frequencies. Electronic charge instabilities within the 
RPA approximation are found by imposing a divergent density-density response, i.e., a vanishing denominator 
$1+V({\qvec_c})\Pi(\qvec_c,0)=0$ first occurring at some finite $\qvec=\qvec_c$. To find the RPA instability line we first adjust 
$\lambda$ and $V_c$ to match the instability point with the doping $p=p^*\approx 0.19$ at which the 
pseudogap crossover line  $T^*(p)$ extracted from resistivity data extrapolates for $T\to 0$. Once 
the parameters of $V(\qvec)$ are fixed, the instability is found at finite $T$ by considering the $T$ 
dependence of the polarization bubble Eq.\,(\ref{bubble}) only. The resulting instability line is given 
by the magenta line in Fig.\,\ref{phase-diag}, remarkably fitting the entire experimental $T^*(p)$ 
line. 

Expanding $V(\qvec)$ and $\Pi(\qvec,\omega_n)$ around $\qvec_c$ and $\omega_n=0$ one obtains the standard 
expression (\ref{fluctuator}) for the quantum-critical charge fluctuation propagator \cite{CDG,enss}
The frequency cutoff $\overline{\Omega}$ is related to the characteristic energy scale of the short-range 
interaction mediators (e.g., for phonons of typical energy $\omega_0$, 
$\overline\Omega\sim\omega_0^2/t\sim \omega_0 /5 \sim 10$\,meV \cite{enss}). Above the mean-field QCP, 
the mass term $m_0\propto \xi^{-2} \sim T^2$ increases, due to reduction of the correlation length $\xi$.

\begin{figure}[htbp]
\vspace{-0.5truecm}
\includegraphics[angle=0,scale=0.15]{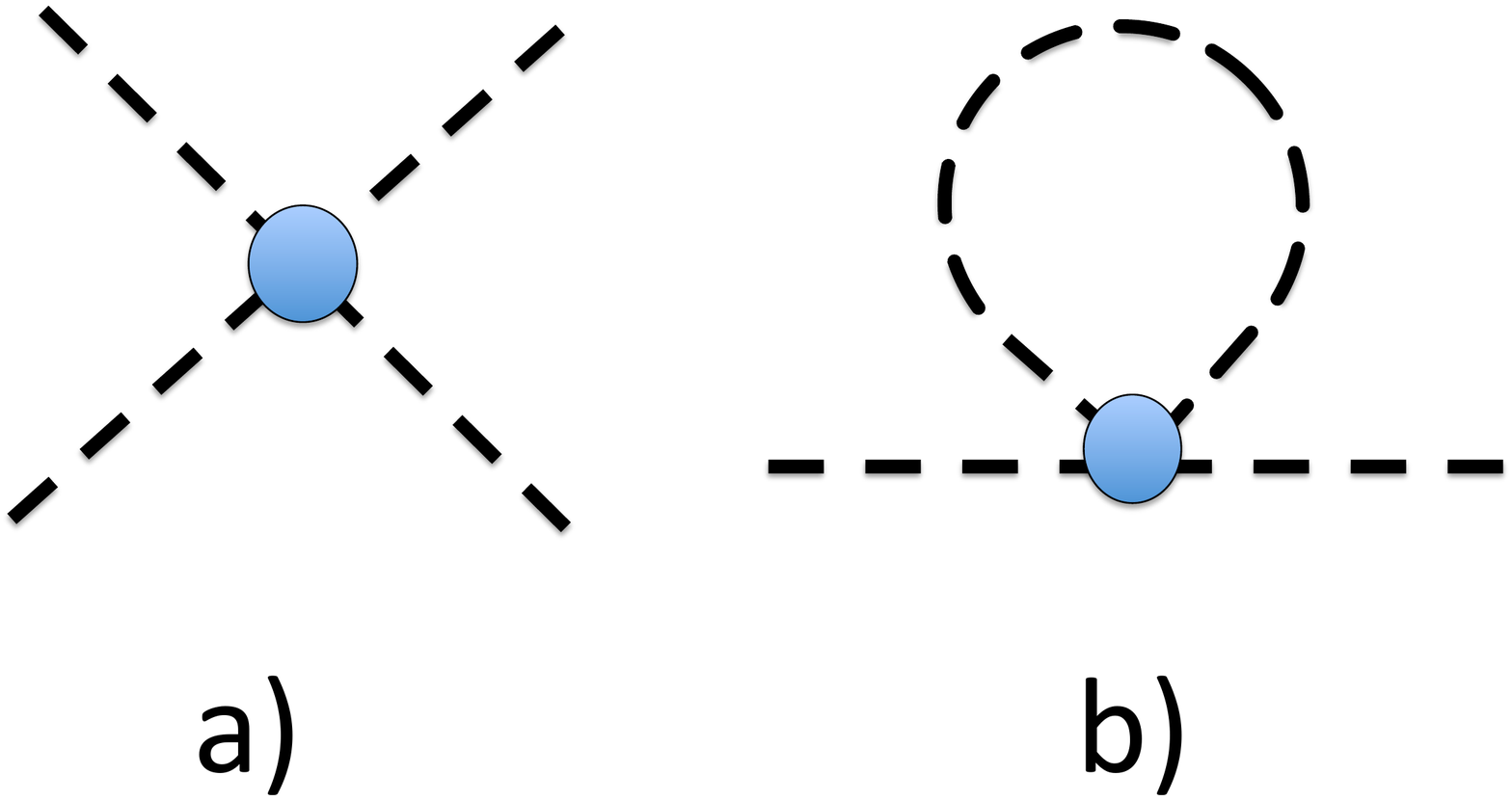}
\caption{ Feynman diagrams for the CDW field. (a) four-leg vertex representing the interaction between CDW fields. (b)
First-order self-energy correction to the CDW fluctuation propagator}
\label{fig-feynman}
\end{figure}

This behavior is modified by the first perturbative correction beyond RPA. Within the standard derivation 
of a Ginzburg-Landau description of criticality in fermion systems \cite{hertz, millis}, this correction 
comes from the $u\Psi^4$ interaction between charge fluctuation fields $\Psi$, which is depicted in 
Fig.\,\ref{fig-feynman}(a). 
The mass correction comes from the contraction of two legs and is represented by the diagram in 
Fig.\,\ref{fig-feynman}(b).

To obtain the blue solid line in Fig.\,\ref{phase-diag} we solved Eq.\,(\ref{renmass}) fixing $\Lambda=1800$\,K and 
$\bar{\Omega}=130$\,K
according to the values obtained fitting Raman spectra in Ref.\,\onlinecite{CDMPHLEKAG}. Then the coupling $u=600$\,K 
(corresponding to a dimensionless coupling $u/\Lambda=1/3$) was adjusted
to match the shifted QCP with the experimental value $p_c\approx 0.16$.
The green line in Fig.\,\ref{phase-diag} is instead obtained by setting $m=\omega_{pr}=50$ K

\section{Hall coefficient within the CDW-QCP model}

The in-plane longitudinal and transverse conductivities have
been derived in Ref.\,\onlinecite{hussey} and read
\begin{eqnarray}
\sigma_{xx}&=&\frac{e^2}{\pi^3\hbar}\frac{2\pi}{d}\int\!\mathrm d\phi\,
\frac{k_F v_F \cos^2(\phi)}{\Gamma(\phi)}\label{eq:rhoxx}\\
\sigma_{xy}&=&\frac{e^3 H}{\pi^3\hbar^2}\frac{2\pi}{d}\int\!\mathrm d\phi\,
\frac{v_F \cos(\phi)}{\Gamma(\phi)}\frac{\partial}{\partial \phi}
\frac{v_F \sin(\phi)}{\Gamma(\phi)} 
\end{eqnarray}
where $\Gamma(\phi)$ is the anisotropic scattering rate along the Fermi surface ($\phi$ denotes the angle 
between the in-plane momentum and its $x-$direction). Following the analysis of Ref.\,\onlinecite{badoux}
we neglect the dependence of Fermi momentum $k_F$ and Fermi velocity $v_F$ on $\phi$ and evaluate these 
quantities from
\begin{eqnarray*}
v_F&=&\frac{\hbar k_F}{m^*}, \\
k_F&=&\frac{\sqrt{2\pi(1+p)}}{a},
\end{eqnarray*}
with hole doping $p$ and an effective mass $m^*=4.1\,m_e$. The in-plane
lattice constant for YBCO is taken as $a=3.85$\,\AA and the distance
between planes is $d=3.2$\,\AA. The Hall coefficient and longitudinal
resistivity are then 
\begin{eqnarray}
R_H&=&\frac{\sigma_{xy}}{\sigma_{xx}^2+\sigma_{xy}^2} \frac{1}{H},\nonumber\\
\rho_{xx}&=&\frac{\sigma_{xx}}{\sigma_{xx}^2+\sigma_{xy}^2}\,.\label{rhoxx}
\end{eqnarray}
  
As discussed in the main text our model for the anisotropic scattering
rate Eq.\,(\ref{gamma1}) is derived from the self-energy for quasiparticles
subject to quantum critical fluctuations in the spin \cite{abanov} or
charge \cite{sulpizi} channel. It is limited by a maximum scattering rate
$\Gamma_{max}=v_F/a$ and impurity scattering is considered via a (doping independent) 
elastic scattering rate which we fix to $\Gamma_0=0.86$\,THz.
The specific form for the quantum critical scattering Eq.\,(\ref{gamma2}) comprises an 
anisotropic and doping dependent mass $M(\phi)=m_0(T)+\bar{\nu} \sin^2(2\phi)$
which is minimized at the hot spots for CDW scattering, i.e., around the
antinodal points $\phi=0, \pi/2$. The parameter $\bar{\nu}=480$\,K is an electronic energy scale 
\cite{enss}, whereas $m_0(T)$ is obtained self-consistently by solving Eq.\,(\ref{renmass}).

\begin{figure}[htbp]
\includegraphics[angle=-0,scale=0.3]{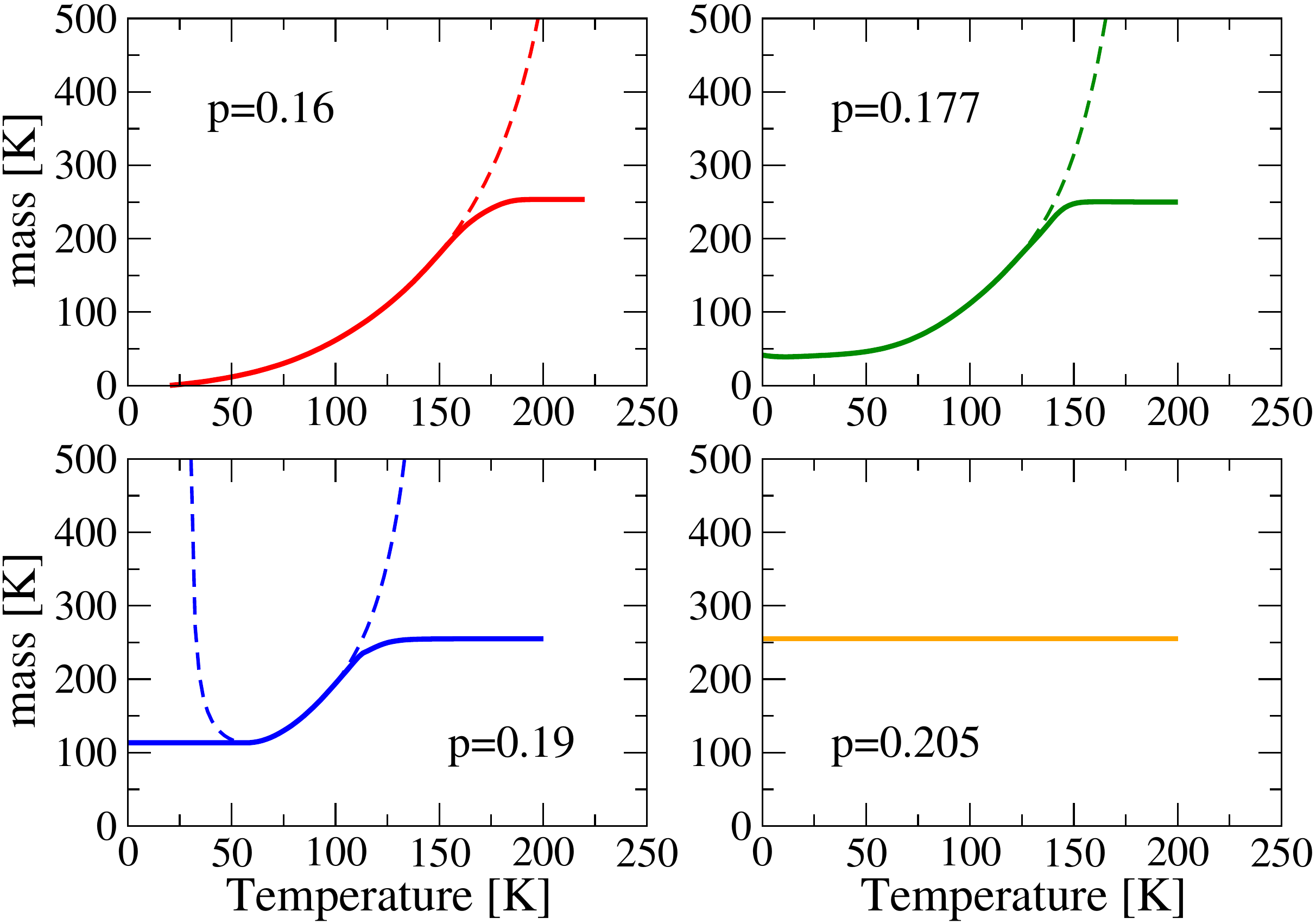}
\caption{Dashed lines: Temperature dependence of the mass term $m_0(T)$ as determined from Eq.\,(\ref{renmass}) 
for the four doping values considered in the paper. Solid lines include a cutoff $m_{max}=255$\,K and 
$m_{min}=112$\,K (for $p=0.19$).}
\label{figMethHall1}
\end{figure}

The dashed lines in Fig.\,\ref{figMethHall1} report the temperature dependent mass for four doping values. 
However, since our theory is only valid in the quantum critical region we cutoff the mass
at a value $m_{max}=255$\,K, which yields the solid lines in Fig.\,\ref{figMethHall1}. Moreover, for doping 
$p=0.19$ we also cutoff the divergence at low temperature which is due to a reentrant behavior caused by 
the nearby van-Hove singularity. We want to stress that these cutoffs do not influence on our main
conclusion, which is drawn from the significant doping dependent change of $R_H$ at $T=50$\,K, but allow us to fit 
$R_H$ over a larger temperature interval as shown in Fig.\,\ref{Hall} of the main text.

\begin{figure}[htbp]
\includegraphics[angle=-0,scale=0.3]{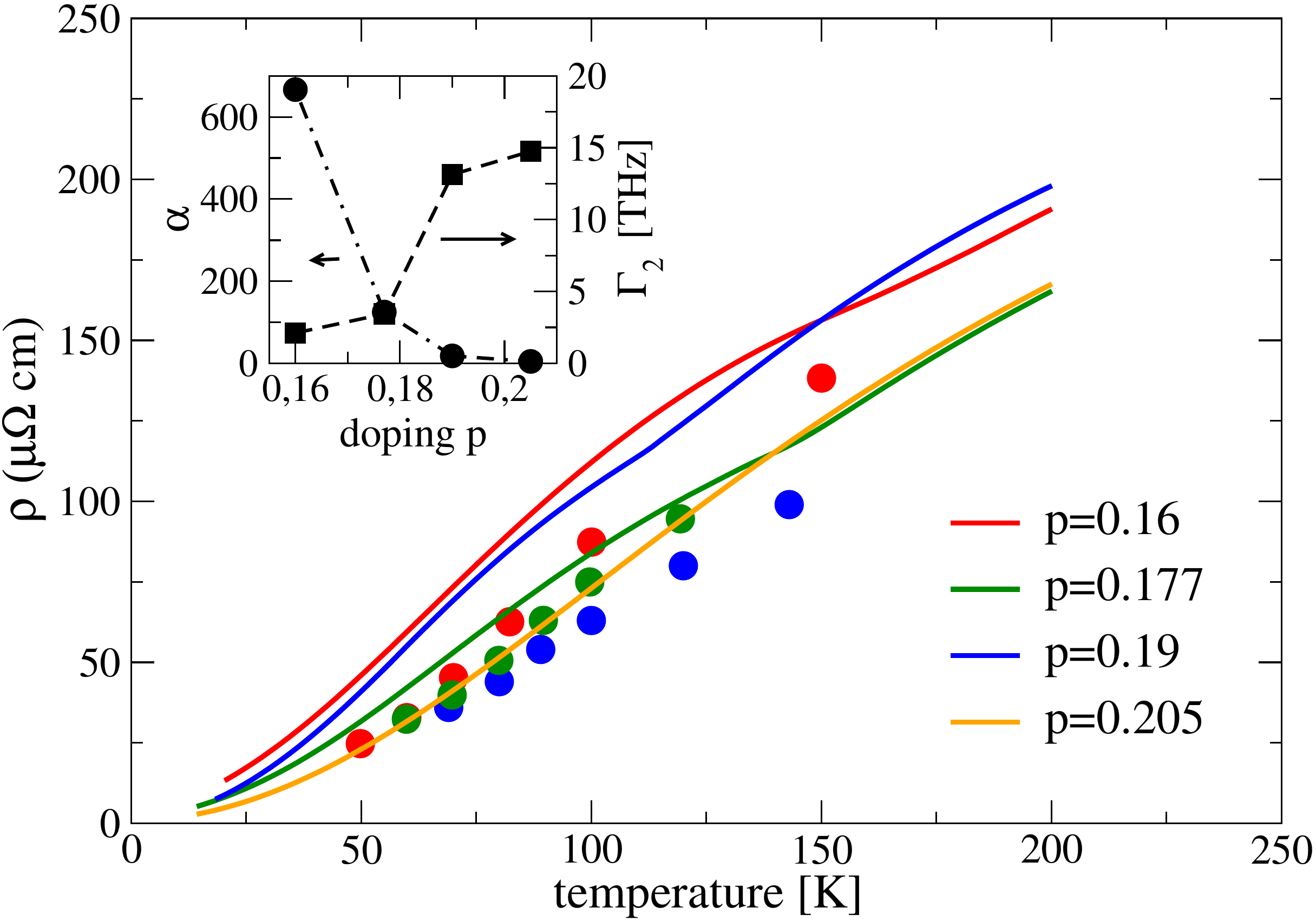}
\caption{Main panel: Fits of the longitudinal resistivity  Eq.\,(\ref{rhoxx})
as compared to the high field normal state data (full dots) from Ref.\,\onlinecite{badoux}. Inset: Doping 
dependence of the fit parameters $\alpha$ (solid circles) and $\Gamma_2$ (solid squares).}
\label{figMethHall2}
\end{figure}

The only fit parameters for the four doping values are the overall
coupling $\Gamma_2$ and the parameter $\alpha$ (cf. Eq.\,\ref{gamma2})
which governs the
anisotropy of the coupling to the quantum critical CDW fluctuations . 
The doping dependence of these parameters is shown in the inset
to Fig.\,\ref{figMethHall2}. The anisotropy parameter $\alpha$ strongly
increases upon approaching the QCP and determines the total coupling
as shown in the inset to Fig.\,\ref{Hall} in the main text PAIRING.
For the same parameter values Fig.\,\ref{figMethHall2} reports the
temperature dependence of the longitudinal conductivity as compared
to the high field normal state data from Ref.\,\onlinecite{badoux}.
It should be noted that the QCP model yields resistivity curves much
closer to the experimental data then the analysis of Ref.\,\onlinecite{badoux}
which was based on a scattering model which anisotropy increases linear
in temperature whereas at higher temperature a Fermi liquid type $T^2$
contribution dominates. In our model the low temperature regime is
usually Fermi liquid like (except at the QCP) whereas the linear
temperature regime is intermediate before crossing over to a $\sqrt{T}$-like
behavior. Moreover, the scattering is anisotropic over the whole
temperature range in contrast to the model in Ref.\,\onlinecite{badoux} where
it is essentially confined to low temperatures.

As discussed in the main text, we disregard here the additional effect of 
pair fluctuations. While this will likely reduce the request of strong anisotropic CDW-mediated scattering
(the parameter $\alpha$) it naturally reduces the value of resistivity improving the agreement between
calculated and experimental resistivity).

\section{CDW suppression by phase fluctuations}

We started from a bare CDW phase stiffness $D_0=A\, p$ linearly increasing with doping as a result of strong correlations
near the Mott-Hubbard transition. In the absence of experimental indications on the value of $A$, we took $A=5800$\,K, which 
fixes $D_0= 580$\,K at $p=0.1$. This choice is rather immaterial in the forthcoming discussion, where we highlight the dependence
of the outcomes of our calculations on quantities that can be accessed by experiments. Of course, any experimental determination
of $D_0$ is accommodated by a change of the other model parameters, so as to keep fixed the measurable quantities.

The reduction of the stiffness in Eq.\,(\ref{phase_fluct}) can be cast in a more convenient way by
using the semiclassical approximation for the Bose function $b(z)\approx T/z$ yielding
$$
D(T)=D_0 \left[1-\frac{u_\theta c_\theta \Lambda_\theta}{6 \pi D_0} \left( 1 +\frac{3T}{2 c_\theta \Lambda_\theta}    \right)\right].
$$
The BKT transition is obtained when $D=2T/\pi$. We point out that, even if 
this transition were not ruled by the two-dimensional BKT physics, Eq.\,(\ref{phase_fluct}) nonetheless implies 
a transition at a temperature $T_\theta$, where the $D$ vanishes, (see blue dotted line in the inset of 
Fig.\,\ref{scheme}). $T_\theta$ is obviously higher than $T_{BKT}$, where $D$ is still finite.

We define the dimensionless quantity $\widetilde{T}=3T/(2c_\theta \Lambda_\theta)$
and the dimensionless bare BKT temperature $\widetilde{T}_0=3T_0/(2c_\theta \Lambda_\theta)$, using the characteristic
energy scale of the phase mode $c_\theta \Lambda_\theta$ as an energy unit. Taking for the sake of convenience 
$\tilde{u}=u_\theta/8$, we obtain from the above equation the condition for the BKT transition temperature suppressed by phase fluctuations, $\widetilde{T}=\widetilde{T}_0 - \tilde{u} \left( 1+ \widetilde{T} \right)$.
Within this approximation, the critical BKT line is then 
$$
\widetilde{T}=\frac{\widetilde{T}_0-\tilde{u}}{1+\tilde{u}}
$$
while from the definition of $\widetilde{T}_0$ and $c_\theta$, we find $\widetilde{T}_0=\beta \sqrt{p}$ with 
$\beta=3\sqrt{A\chi}/(2\Lambda_\theta)$. The condition for a QCP where the BKT transition temperature vanishes is 
then given by  $\widetilde{T}_0=\tilde{u}$, which also defines the critical doping $p'_c\approx 0.08$ fixed by 
experiments \cite{badoux}. Expanding around this value, we find the equation for the critical line
$$
\widetilde{T}=\frac{\tilde{u}}{2(1+\tilde{u})\sqrt{p'_c}}\left(p-p'_c\right)
$$
The slope of  the physical $T_{CDW}(p)$ line is obtained by reintroducing the characteristic energy scale 
$c_\theta \Lambda_\theta$ 
$$
\frac{c_\theta \Lambda_\theta \tilde{u}}{2(1+\tilde{u})\sqrt{p'_c}}
$$
and is fixed by fitting the data (light-blue squares in Fig.\,\ref{phase-diag}). In this way, the 
various physically relevant combinations of the microscopic parameters have been reduced to two effective parameters 
only, the position of the QCP and the slope of the $T_{CDW}(p)$ curve, which can be fixed by experiments. Specifically 
we obtain the dashed blue line in Fig.\,\ref{phase-diag} with $\tilde{u}=4.5$, and the characteristic energy scale 
$c_\theta= 850 \sqrt{p}$\,K and numerically solving the full expression Eq.\,(\ref{phase_fluct}).


\end{document}